\documentclass[prd,preprint,nofootinbib,showpacs]{revtex4}
\usepackage{amsmath,amssymb,graphicx,subfigure,bm}

\newcommand{\be}{\begin{equation}}
\newcommand{\ee}{\end{equation}}
\newcommand{\bea}{\begin{eqnarray}}
\newcommand{\eea}{\end{eqnarray}}
\newcommand{\N}{\mathcal{N}}
\renewcommand{\b}[1]{\bar{#1}}
\newcommand{\Del}{\nabla}
\newcommand{\del}{\partial}

\newcommand{\bj}{\bar{\jmath}}
\renewcommand{\ap}{\alpha^\prime}

\newcommand{\im}{\mathnormal{Im}}
\newcommand{\comment}[1]{}
\newcommand{\bd}{\ensuremath{\overline{\textnormal{D3}}}}
\renewcommand{\k}{\kappa_4}

\newcommand{\gs}{g_s}

\newcommand{\gt}{\tilde{G}}

\newcommand{\rt}{\tilde{r}}
\newcommand{\fL}{\mathcal{L}}

\newcommand{\rds}{\dot{r}^2}

\newcommand{\td}{\dot{t}}

\newcommand{\tds}{\dot{t}^2}

\newcommand{\hi}{h^{-1}}

\newcommand{\tmhr}{\tds - h \rds}

\newcommand{\cf}{\tilde{C}_4}
\newcommand{\wo}{W_0}

\begin{document}
\preprint{hep-th/0305018}

\title{The Fall of Stringy de Sitter}

\author{Andrew R. Frey}
\email{frey@vulcan.physics.ucsb.edu}

\author{Matthew Lippert}
\email{lippert@physics.ucsb.edu}

\author{Brook Williams}
\email{brook@physics.ucsb.edu}

\affiliation{Department of Physics\\
University of California\\ 
Santa Barbara, CA 93106, USA}


\begin{abstract}
Kachru, Kallosh, Linde, \& Trivedi recently constructed a four-dimensional
de Sitter compactification of IIB string theory, which they showed to be
metastable in agreement with general arguments about de Sitter spacetimes
in quantum gravity.  In this paper, we describe how discrete flux choices
lead to a closely-spaced set of vacua and explore various decay channels.  
We find that in many situations NS5-brane meditated decays which
exchange NSNS 3-form flux for D3-branes are comparatively very fast. 

\end{abstract}

\date{\today}

\maketitle

\section{Instabilities of dS Flux Compactifications}\label{s:intro}

De Sitter spacetime (dS) holds a special place in the study of quantum
gravity.  Constructing and exploring the maximally symmetric spacetime
with positive cosmological constant $\Lambda$ has been the source of
much recent interest despite (or perhaps because of) its stubborn
opacity. While much progress has been made in the understanding of the
other maximally symmetric solutions, Minkowski and anti-de Sitter
(AdS) spaces, dS has until recently eluded string theoretic
description because of some of its unique properties.  The
observer-dependent horizon of dS, like a black hole horizon, yields a
thermal state with finite entropy.  Not only are the S-matrix
observables of string theory  precluded in $\Lambda > 0$ spaces
\cite{Hellerman:2001yi}, but, due to  the inevitable Poincar\'e
recurrences \cite{Dyson:2002pf}, all observables  are ill-defined
\cite{Banks:2002wr}.  These issues would merely be of abstract
theoretical importance were it not for recent observational evidence
\cite{Bennett:2003bz} indicating that not only was $\Lambda > 0 $
during in the early universe during inflation, but it seems to be so
today.

It has become increasingly clear that dS cannot be a stable state in
any theory of quantum gravity.  The symmetries of dS are incommensurate with 
the discrete spectrum implied by finite entropy \cite{Goheer:2002vf}.  
Rather than a stable
vacuum, dS is instead a metastable resonance whose lifetime, on
general entropic grounds, must be less than the recurrence time
\cite{Kachru:2003aw,Goheer:2002vf,Giddings:2003zw}.

One would ask, then, what does string theory say about dS and its
decay  modes?  String models of dS have been difficult to find partly
because, as non-supersymmetric vacua, they are isolated points in
moduli space with all moduli stabilized. Notably, some dS
compactifications of string theory were described in
\cite{Silverstein:2001xn,Berglund:2001aj} and, in a well-controlled
manner for critical strings, by Kachru, Kallosh, Linde, \& Trivedi
(KKLT) \cite{Kachru:2003aw}. Generically, any string theoretic dS
compactification can decay and decompactify \cite{Dine:1985he,Giddings:2003zw}
because the 10D Poincar\'e invariant string vacuum is
supersymmetric and so has vanishing energy density.  However, this is
far from the only decay mode.  For example, in any compactification
in which RR fluxes contribute to the potential, D-brane instantons
change the fluxes and the cosmological constant.  This has been an
object of study in many papers, including
\cite{Brown:1987dd,Brown:1988kg,Bousso:2000xa,Feng:2000if,Maloney:2002rr}.

We consider a slight twist on the brane instanton decays.  
In \cite{Kachru:2003aw}, the cosmological constant gets a positive 
contribution from \bd-branes, rather than directly from the fluxes.  
This effect has been seen in the AdS/CFT correspondence, where instantonic
NS5-branes provide a decay mode for the \bd-branes \cite{Kachru:2002gs}.
These results apply to the similar dS compactifications of 
KKLT and are particularly of interest because they can 
end in a state of positive cosmological constant.  Therefore,
one might wonder whether this type of decay could occur quickly enough
to affect the cosmological constant within the age of the universe.  In this 
paper, we generalize the results of Kachru, Pearson, \& Verlinde (KPV) 
\cite{Kachru:2002gs} to dS 
compactifications and compare the decay rate through the 5-brane channel to
two other decays, one to decompactification and the other by \bd-brane
tunneling in the compactification manifold.  
We give explicit examples in which the 5-brane decays are much faster
than the others.

In the next section, we review the dS vacuum construction that we will
study.  In section \ref{s:parameters}, we flesh out the discrete
landscape of vacua that are available through tuning and among which
our instantons will interpolate.  We then review the AdS/CFT
instantons of  \cite{Kachru:2002gs} and make the corrections
necessary to compactify their backgrounds in section
\ref{s:kpvinst}.  We apply our calculation to find decay
times for specific sets of initial parameters in section \ref{s:apply} and compare them to
those of KKLT in section \ref{s:compare}.  In addition, we comment on two
other possible decay channels.  We will
generally keep factors of the gravitational coupling $\k$ and the
string length $\ap$ explicit in formulae, but any numbers we cite
should be taken in Planck/string units.

\section{Building dS Vacua}\label{s:kklt}

Constructing a solution of string or M-theory with a four-dimensional
dS vacuum  has been a longstanding challenge.  Such a solution must
be non-supersymmetric  and requires aspects of the theory beyond the
low-energy SUGRA limit.

Recently, however, KKLT \cite{Kachru:2003aw} presented 
a specific construction in critical string theory with no unfixed
moduli.  The model was based on the warped flux compactifications
studied by Giddings, Kachru, \& Polchinski (GKP)
\cite{Giddings:2001yu}\footnote{The  GKP type of compactification was
studied earlier in simpler cases and in  M-theory by
\cite{Gukov:1999ya,Dasgupta:1999ss,Greene:2000gh,Chan:2000ms}.  
The  supersymmetry
conditions and equations of motion were considered in
\cite{Becker:1996gj,Grana:2000jj,Gubser:2000vg}.  Explicit constructions
on tori and K3 are in \cite{Kachru:2002he,Frey:2002hf,Tripathy:2002qw}.}.  
Non-perturbative
corrections fix the overall K\"{a}hler modulus of this tree-level
no-scale model, resulting in a stable, supersymmetric AdS vacuum.
KKLT then added \bd-branes to yield a metastable dS vacuum and showed,
by considering decays to decompactification, the lifetime to be less
than the Poincar\'{e} recurrence time.

The GKP compactification of IIB string theory on a 
threefold $M$ with
7-branes and O3-planes can be efficiently described as an F-theory 
compactification
on a CY fourfold (CY) $X$.   $X$ is elliptically fibered over
$M$ such that the fiber's complex structure $\tau = c_0 + i e^{-\phi}$
is the IIB axion-dilaton (we take for simplicity $\tau=i/g_s$).
We will consider the orientifold limit of F-theory in which $M$ is an 
orientifolded CY threefold.
Three-form fluxes and D3-branes are added subject to the global
tadpole constraint, or the global conservation of RR 5-form $F_5$
flux:  
\be
\label{tadpole} 
0 = N_{\textnormal{D3}} - N_{\bd} + \frac{1}{2\kappa_{10}^2 \mu_3} 
\int_M H_3 \wedge  F_3 - \frac{\chi (X)}{24}\ . 
\ee
The Euler number of
the CY fourfold $\chi(X)$ gives the effective negative D3-brane charge
in IIB of O3-planes and D7-branes wrapped  on 4-cycles of $M$.  For typical
choices of $X$, $\chi(X)$ can be up to $O(10^5) $
\cite{Klemm:1998ts}. This is must be balanced by the charge from
4D space-filling D3-branes, \bd-branes, and the wrapped NSNS
and RR 3-form fluxes $H_3$ and $F_3$, which also source $F_5$.

To construct their model, KKLT began by choosing $X$ and a set of
wrapped fluxes, while setting $N_{\textnormal{D3}} = N_{\bd} = 0$.  
The CY threefold $M$ has
$b_3 \gg 1$ three-cycles,  and a particular choice fluxes  $H_3$, $F_3 \in
H^3(M,Z)$ represents a point in a $2b_3$ dimensional  lattice.  The
fluxes combine into a single complex 3-form $G_3 = F_3 - \tau  H_3$.
For simplicity, KKLT chose $h^{1,1}(X)= 2$, so that $M$ has a single
K\"ahler modulus $\rho$. In addition to the moduli $\tau$ and $\rho$,
$M$ has $h^{2,1}(M)$ complex structure moduli $z^\alpha$.

In the presence of fluxes, the classical 4D effective $\N=1$
superpotential is \cite{Gukov:1999ya} \be
\label{superpotential}
\wo = \frac{1}{\k^8}\int_M G_3 \wedge \Omega\ , 
\ee
where $\Omega$ is the holomorphic (3,0) form on $M$.  $\wo$ then 
is given by the (0,3) part of
the $G_3$ flux which, because the fluxes are quantized, can only be
tuned discretely.  The tree-level K\"ahler potential (ignoring warping
of the  spacetime metric) 
\be
\label{kahlerkklt}
\mathcal{K} = -3
\log(-i(\rho-\b\rho)) - \log(-i(\tau-\b\tau)) - \log\left(-\frac{i}{\k^6}  
\int_M \Omega \wedge \b\Omega\right) 
\ee 
along with $\wo$ gives the no-scale potential
\be\label{noscale}
V = e^{\mathcal{K}} \sum_{i,\bj} \mathcal{K}^{i\bj} D_iW \b D_{\bj} \b W
\ee
where $i, j$ sum over all moduli but $\rho$,  $\mathcal{K}_{i\bj}
= \del_i \del_{\bj} \mathcal{K}$ is the K\"ahler metric, and $D_i =
\del_i +\del_i \mathcal{K}$ is the  K\"ahler covariant derivative.
Except for the volume modulus $\rho$,  this potential generically
fixes all other moduli  such that $G_3$ is imaginary
self-dual.\footnote{It is natural to wonder if corrections to the
K\"ahler potential due to warping could fix the radial modulus.  While
the precise form of $\mathcal{K}$ is difficult to compute for $\rho$,
because the 10D solution exists at tree level  for all
compactification scales, the final potential must be no-scale
\cite{DeWolfe:2002nn}  We will look at warping in the complex
structure K\"ahler potential below.}   The remaining condition for
supersymmetry,  $D_\rho W= 0$ is satisfied only when $\wo = 0$, which
implies  that in supersymmetric vacua $G_3$ is a (2,1) form.

The geometry of $M$ is, of course, very complicated but is accurately
described  near conifold points by the Klebanov-Strassler (KS)
solution \cite{Klebanov:2000hb}.   Wrapped  fluxes warp and deform the
conifold; at the tip $y=0$,  the metric is
\be
\label{tipmetric}
ds^2 = h^{-1/2} \eta_{\mu\nu} dx^\mu dx^\nu +b g_s M\ap \left(
e^{2u}dy^2 +d\Omega_3^2+e^{2u}y^2 d\Omega_2^2\right)
\ee
 where $b\sim 1$ is a numerical constant and $e^{u}$ is the
compactification length scale (here we use 10D string frame).   Notice
that the $S^3$ at the tip has a fixed proper size depending only on
the fluxes.  Also, the $S^2$ is nontrivially fibered over the $S^3$.  
Away from the tip, the throat has approximately a warped
conifold metric  
\be 
\label{conifold} ds^2\approx h^{-1/2} \eta_{\mu
\nu} dx^\mu dx^\nu + h^{1/2} e^{2u}(dr^2 + r^2 ds^2_{T^{1,1}}) 
\ee
where $ds^2_{T^{1,1}}$ is the metric on the base $T^{1,1}$.  In this region,
the warp factor is approximately
\be\label{throatwarp} 
h = 1 + (L^4/r^4) \log(r/r_s)\ee 
with the length scale $L^4 = \frac{81}{8}e^{-4u}g_s M \ap$.  Here, the radial coordinates
$r$ and $y$ are complicated functions of each other, and the tip is at
$y=0, r=\rt$.  For the undeformed conifold, the singular tip is
located at  $r=r_s=\rt e^{-1/4}$.  Splitting the conifold into
the  tip and throat in this manner is described in
\cite{Herzog:2001xk} and references therein. 
                                                                       
The radial modulus $\im\, \rho = e^{4u}/g_s\equiv\sigma$  is defined
so that, at large radius, the total unwarped  volume of the
compactification is $\int_M d^6x \sqrt{h^{-1/2}g} \approx
e^{6u}\ap{}^3$.  The fluxes through any 3-cycle of $M$ are quantized,
and for a given conifold throat 
\be 
M = \frac{1}{4\pi^2 \ap} \int_A F_3 \ ,\ \ K =  -\frac{1}{4\pi^2 \ap} 
\int_B H_3\label{fluxquant} 
\ee
where the $A$ cycle is the $S^3$ which stays finite at the tip and the
$B$ cycle is the six-dimensional dual of $A$.   GKP found that the
warp factor at the tip of the conifold is related to the deformation
parameter $z$ of the tip,  which is determined by the flux
superpotential (\ref{superpotential}), by 
\be\label{tipwarp} h(y
=0) \approx \frac{(g_s M)^2}{|z|^{4/3}}\ ,\ \ z= \exp \left[- \frac{2\pi
K}{g_s M}\right]\ .  \ee 
It is this particular form of the warp factor
that gives the $A$ cycle a fixed proper size at the tip.  Note that
this is not the $r \to \rt$ limit of equation (\ref{throatwarp}) because the conifold is deformed.

To generate a nontrivial potential for $\rho$, as suggested in 
\cite{Giddings:2001yu}, KKLT considered
non-perturbative  corrections to the superpotential
(\ref{superpotential}).   Both wrapped Euclidean D3-branes and gluino
condensation on the worldvolume of non-Abelian D7-branes  generate
additional terms of the form \be\label{nonperturb} \delta W = A
e^{ia\rho} \ee where the constants $A \sim O(1)$ and $a \sim
O(10^{-1})$.   For simplicity, KKLT took $\rho$ to be purely
imaginary,  $\rho = i \sigma$, and $A$, $a$, $\wo$  to be real.  The
potential now becomes 
\be
\label{nonpertV} 
V = \frac{aAe^{-a\sigma}}{2\sigma^2} \lbrace  Ae^{-a\sigma}  (1+
\frac{a\sigma}{3} ) + \wo \rbrace, 
\ee
 and for suitable $\wo < 0$
there is a supersymmetric vacuum with $V_0<0$, implying the
non-compact directions are AdS.  For $|\wo| \ll 1$, the AdS
minimum lies at $\sigma_{cr} \gg 1$ where the SUGRA can be trusted and
$\ap$  corrections are small.

The final step in the KKLT construction is to add enough \bd-branes so
that $V_0 > 0$ and the vacuum is dS.  The global $F_5$ charge must
still be conserved via eqn (\ref{tadpole}), and the addition of $p$
\bd-branes gives $N_{D3} = -p$.  By adjusting the fluxes, a
corresponding increase in $\int_M H_3 \wedge F_3$ balances this
reduction.  The \bd-branes break supersymmetry and add some extra
energy \cite{Kachru:2002gs}, \be\label{deltaV} \delta V = \frac{D
p}{\sigma^3}; \ \ \ D = 2 \mu_3 h^{-1}(r)  \ee   where $\mu_3$ is the
brane charge.  To minimize their energy, the \bd-branes migrate to a
conifold tip, so the energy density per \bd-brane depends,  through
eqn (\ref{tipwarp}), on the fluxes.  For sufficiently fine-tuned
parameters, this additional term in the potential lifts the AdS global
minimum to a dS local minimum.

Unlike the AdS vacuum, the dS minimum is only metastable.  KKLT
investigated one  possible decay mode, tunneling to large $\sigma$.
The potential becomes arbitrarily close to zero at large radius, so it
is possible to tunnel to a runaway, decompactifying solution.

Coleman and De Luccia (CDL) \cite{Coleman:1980aw} described such an
instanton including gravitational back-reaction.  In terms of a
canonical scalar  field $\varphi =(\sqrt{3/2}\log\sigma)/\k$, the
Euclidean action is 
\bea S_E[\varphi] &=& \int d^4x \sqrt{g}
(-\frac{1}{2\k^2}R + \frac{1}{2}(\del  \varphi)^2 + V(\varphi))
\nonumber\\ &=& -\int d^4x \sqrt{g} V(\varphi)\label{CDLaction} 
\eea
using Einstein's equations to get the second line.  The instanton
$\varphi_{CDL}$ is an O(4)-symmetric interpolation between the dS
vacuum at  $\varphi_{cr}$ and the supersymmetric vacuum at $\varphi =
\infty$.  When Wick rotated back, this gives the usual expanding
bubble of true vacuum inside the false dS vacuum.   The action of the
static dS vacuum is simply computed to give
\be\label{backgroundaction} S_0= -\frac{24 \pi^2}{\k^4 V_0} =
-\bm{S_0} \ee where $\bm{S_0}$ is the entropy of the dS vacuum.  The
tunneling probability per unit volume
is given by the difference between the action
of the instanton solution  and the static dS vacuum:
\be\label{cdlprob1} P^{CDL}_{decay} \sim e^{-S[\varphi_{CDL}] + S_0} \ .  
\ee
From eqn (\ref{CDLaction}), $S[\varphi] < 0$ for $V(\varphi) > 0$,
and the  resulting lifetime is exponentially less than the Poincar\'{e}
recurrence  time $t_r \sim e^{\bm{S_0}}$: 
\be\label{cdltime} 
t^{CDL}_{decay} \sim e^{\bm{S_0} - |S[\varphi] |} < t_r 
\ee 
which is in line with the general arguments of
\cite{Goheer:2002vf,Susskind:2003kw,Giddings:2003zw}.

In addition to the CDL instanton, KKLT considered decompactification
decay via the stochastic Hawking-Moss (HM) instanton
\cite{Hawking:1982fz}.  Considering decays of general dS string
compactifications, \cite{Goheer:2002vf} and \cite{Giddings:2003zw}
also discussed thermal fluctuations using the HM reasoning.  Whereas
the CDL instanton tunnels through the potential barrier, the HM
instanton relies on thermal fluctuations to carry $\varphi$ to the top
of the potential, where it can then roll down the other side to the
true vacuum.  While the original HM process is homogeneous, KKLT
argued it should be interpreted as a horizon-sized fluctuation.  If
the potential has a broad, flat maximum at $\varphi_1$, the state
there is approximately dS with energy $V(\varphi_1) > V_0$ and entropy
$\bm{S_1}$.  The probability per unit volume for a thermal fluctuation
is given by the
difference in entropies between the fluctuation and equilibrium:
\be\label{HMprob}
P^{HM}_{decay} \sim e^{\bm{S_1} - \bm{S_0}}  \ . 
\ee
The decay time $t^{HM}_{decay} = (P^{HM}_{decay})^{-1}$ is again less than the recurrence
time $t_r$ and is also less than the CDL decay time (\ref{cdltime})
when the potential barrier is short and wide and thus the thin-wall
approximation is invalid.

%
%
\section{Finding dS Parameters}\label{s:parameters}

As described in section \ref{s:kklt}, obtaining the vacua constructed
in \cite{Kachru:2003aw} requires fine tuning subject to several
constraints.  First, one must adjust the bulk fluxes so that $|\wo|
\ll 1$.   Moreover, a dS minimum requires fine tuning of the fluxes,
$K$ and $M$,  in the KS throat.  A given value of $\wo$ tightly
constrains one's choice for $D p$ (cf. eqn \ref{deltaV}).  
For example, KKLT presented a
model with $\wo = -10^{-4}$ and an AdS minimum of $V_0=-2.00\times 10^{-15}$;
by adding one \bd-brane with $D = 3 \times 10^{-9}$, they achieved
a dS minimum of $V_0=1.77\times 10^{-17}$.  This is a
very special choice of fluxes indeed.  For $D p \lesssim 3 \times
10^{-9}$ the minimum is  at $V_0 < 0$ and is AdS, and for $D p \gtrsim
7.5 \times 10^{-9}$ a local minimum no longer exists.  There are
additional constraints as well.  In \cite{Kachru:2002gs}  it is shown
that there exists a classical instability if  $p/M \gtrsim 0.08$.
Furthermore, results from section 3 of \cite{Giddings:2001yu} rely on
approximations valid when  $K/(g_s M) \gtrsim 1/2$.

\begin{figure}
\begin{center}
\includegraphics{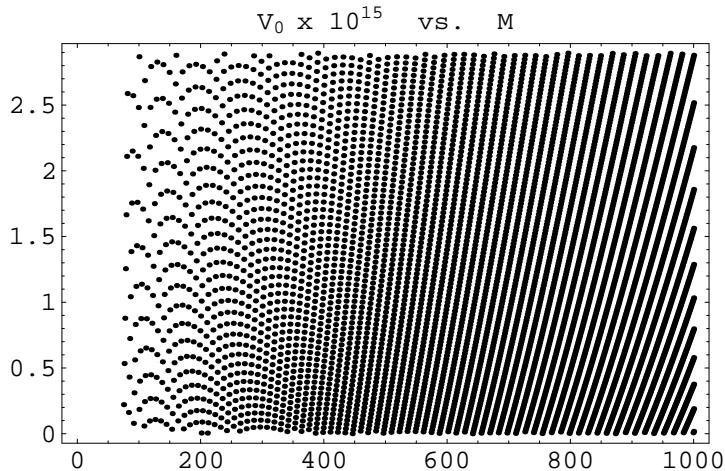}
\end{center}
\caption{The possible dS vacua with $V_0$ for given $M$ illustrate the
density of states consistent with a discretuum.}
\label{f:disc1}
\end{figure}

With such fine tuning and taking into account that the tuning
parameters $K$ and $M$ are discrete, one might question if it is
possible to build such a model at all. Such tuning would require the
existence of a ``discretuum''\footnote{The authors of
\cite{Bousso:2000xa} coined this term to refer to situations in which
a discrete spectrum is sufficiently dense to allow for an (almost)
arbitrarily fine tuning.  Our discretuum is not as finely spaced as
those in \cite{Bousso:2000xa}.}.  We have done numerical searches in
order to map out the discrete landscape of dS vacua.  Figure
\ref{f:disc1} shows the existence of the discretuum.
%
%
Here we have plotted the possible values $V_0$ that have a dS minimum
and can be achieved with integer fluxes for the parameters used in
KKLT,  $\wo = -10^{-4},\  a = .1,\  A = 1,\ g_s=0.1,\ \k = \ap =1$.\footnote{In addition to tuning $V_0$ by varying the fluxes $M$ and $K$, one could, in principle, vary $\wo$ by adjusting the bulk fluxes.  While this would certainly increase the discretuum density, we leave $\wo$ constant as explicit calculation of $\wo$ in terms of bulk fluxes is prohibitively complicated.}  It is clear
that for a desired value of $V_0$ there exists a configuration of
fluxes  with $\Tilde{V}_0 = V_0 + \epsilon$, where $\epsilon$ is very
small; i.e.\ a discretuum does exist!  For each of the models studied
$K/g_s M > 1/2$.  Here we have allowed $M$ to
range from 75 to 1000.  The lower bound avoids the classical
instability (for $p \leq 6$)  As one goes to higher and higher values
of $M$, one must also increase the amount of induced D3-brane charge on
the D7-branes in order to satisfy (\ref{tadpole}).  This might require
adding more D7-branes and, thus, more degrees of freedom, which,
though massive, could cause problems when considering loop
corrections.  

The smallest possible value of $V_0$, for the parameters used in KKLT,
is $\mathcal{O}(10^{-20})$, a far cry from the desired
$\mathcal{O}(10^{-120})$.  In order to obtain a more realistic vacuum
energy, one must attempt to construct a background
with $|\wo| \sim \mathcal{O}(10^{-55})$.  While such a fine tuning seems improbable, with $b_3$ sufficiently large, it is at least possible\footnote{One can estimate the smallest $|\wo|$ to have $\log(|\wo|) \sim -2b_3$.  We thank S. Kachru for discussion on this point.}, if not particularly natural \cite{Bousso:2000xa}.

We have so far considered only a single KS throat, as in KKLT.
However, a general CY has many of them. By considering backgrounds
with multiple KS throats the discretuum density is increased 
dramatically.  One finds that (\ref{deltaV}) becomes,
\be\label{MdeltaV} 
\delta V = \sum_i \frac{D_i \ p_i}{\sigma^3}; 
\ \ \ D_i = 2\mu_3 h^{-1}(\rt_i) \ .  
\ee
where $i$ labels the different
throats.  Clearly, by adjusting the fluxes in each individual throat,
one may tune $\delta V$ with greater accuracy.  
For a single KS throat we found $\mathcal{O}(10^3)$ 
configurations with a dS minimum.
Analogously, for 2 KS throats $(75 \leq M_1 \leq M_2,\ 75 \leq M_2
\leq 300)$ we find $\mathcal{O}(10^5)$ dS minima.   It is
easy to find configurations with $\mathcal{O}(10)$  KS
throats\footnote{For example, in \cite{Greene:1996cy} a family of
quintics are constructed with 16 conifold singularities.}, leading to
an amazingly dense set of vacua.  The inclusion of a second throat
also lowers our minimum value  of $V_0$ by an order of magnitude.
Though this is nice, it does little good in helping build a model with
a realistic cosmological constant.  We suspect that even with the
addition of 10 or more throats the lofty goal of $V_0 \sim 10^{-120}$
would still be far out  of reach.

The following sections describe various decays, analogous to those
studied in \cite{Kachru:2002gs}, in which one unit of $H_3$ flux is
exchanged for $M$ D3-branes.  For geometries with single KS throats, 
after one decay the final state has a negative
cosmological constant and a big crunch in its future.   It has been
argued that these decays should not be  allowed in a quantum theory of
gravity and also that instantons mediating these decays may not be
possible to construct \cite{Banks:2002nm}.  We will not worry about
these subtleties (other than the well-known effects on the instanton
action \cite{Coleman:1980aw}) since our main focus is on 
instanton decays ending in dS.  The
configuration with multiple KS throats is more interesting.  As with
the single throat, these may decay directly into states with negative
cosmological constant.  However, there can now be decays from one dS
vacuum to another with smaller $\Lambda$ (modulo some classical evolution
we will discuss later).  This process is of
particular interest, since it allows for a rather generic set of
fluxes on several KS throats to undergo a series of decays to dS vacua
with smaller and smaller cosmological constant; this situation is
similar to that envisioned by \cite{Abbott:1985qf} and expanded upon
by  \cite{Brown:1987dd,Brown:1988kg,Feng:2000if}.

\section{Decays \textit{\`a la} KPV}\label{s:kpvinst}

\subsection{Review of NS5-brane Instantons}\label{ss:reviewkpv}

The KS geometry found at conifold points of GKP compactifications was
first studied in the usual decoupling limit of string-gauge theory
dualities \cite{Klebanov:2000hb}.  The relevant gauge theory dual is a
duality cascade with an energy dependent effective number of
D3-branes; in the IR, most  of the D3-branes have been transformed
into 3-form fluxes.  The BPS domain wall that transforms the D3-branes
to fluxes was described by KPV; it is a polarized NS5-brane that
carries D3-brane charges and bends over the $A$ cycle at the deformed
conifold tip \cite{Kachru:2002gs}.  As the NS5-brane moves over the
$A$ cycle, the D3-branes are absorbed into the background RR flux, and
the background NSNS flux jumps by a unit due to the NS5-brane charge.

\begin{figure}
\begin{center}
\includegraphics[scale=0.5]{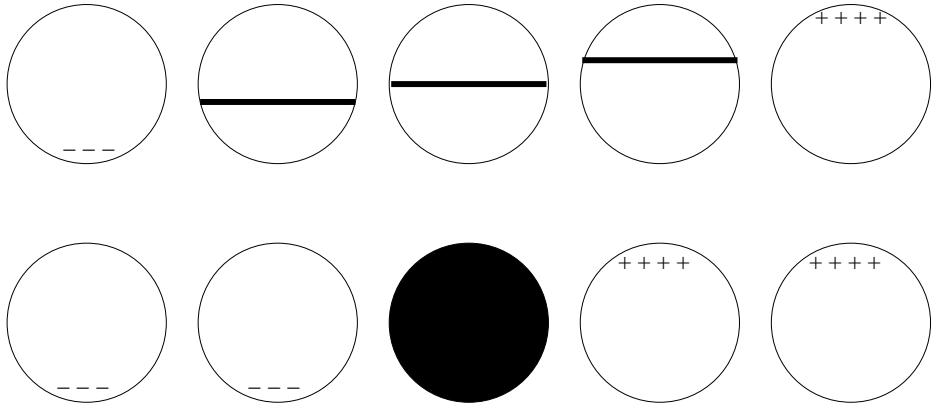}
\end{center}
\caption{\textbf{Top:} In the KPV process, $p$ \bd-branes polarize
into an NS5-brane wrapping an $S^2$ on the $A$ cycle.  The NS5-brane
then slides to the opposite pole, becoming $M-p$
D3-branes. \textbf{Bottom:} In the thin-wall limit, the NS5-brane is
instead wraps the $A$ cycle at a particular Euclidean radius.}
\label{f:thinwall}
\end{figure}

KPV also described nonsupersymmetric gauge theories with $p$
\bd-branes at the tip of the conifold, as in KKLT.  Due to the
3-form flux background, the \bd-branes suffer a classical instability
to brane polarization (first discovered in \cite{Myers:1999ps}) as an
NS5-brane wrapping an $S^2$  in the $A$ cycle.  However, for $p\gtrsim
M/12$, the NS5-brane itself is unstable to collapse around the $A$
cycle, reducing the NSNS flux and turning the \bd-branes into
supersymmetric D3-branes.  For smaller $p$, the decay of the NS5-brane
proceeds by tunneling; in Euclidean spacetime, the NS5-brane is
slightly polarized with \bd\ charge at infinity and bends around the
$A$ cycle to leave D3-branes at the origin \cite{Kachru:2002gs}.  This
process is illustrated in the top line of figure \ref{f:thinwall}.

For $p$ small enough, KPV showed numerically that the thin-wall
approximation is very reasonable.  In that limit, the instanton
appears to be an NS5-brane wrapping the full $S^3$ of the $A$ cycle at
a fixed radius, as shown in the bottom line of figure
\ref{f:thinwall}.  The wrapped $F_3$ flux induces $M$ units of charge
in the NS5 worldvolume gauge theory which is canceled by the charge
carried by the ends of $M$ D3-branes.  The $p$ \bd-branes end on the
outside of the NS5-brane, and $M-p$ D3-branes end on the inside.  The
bubble tension in the effective theory  is just the NS5-brane tension
times the volume of the $A$ cycle. These instantons are clearly
related to the BPS domain walls KPV found.
%
%

In the rest of this paper, we will focus on the thin-wall limit to
estimate the instanton bubble tension.  As has been argued strenuously
\cite{Banks:2002nm}, the thin-wall limit certainly does not describe
the full picture of the decays, but the other contributions to the
Euclidean path integral (such as Hawking-Moss instantons at the other
extreme)  should only enhance the decay rate.  Therefore, we take the
point of view that the thin-wall limit estimates an upper limit for
the decay time.  As a consequence of the thin-wall limit, we may, as
in  KKLT, ignore the polarization of the \bd-branes in the initial
metastable state.  Before we turn to the modifications necessary for
including KPV instantons in KKLT compactifications, let us also note
that our instantons are cousins of the supersymmetry-changing domain
wall bubbles found in \cite{Kachru:2002ns}, just as the AdS/CFT
instantons of KPV  are related to the BPS domain walls.

\subsection{Corrections for Compactifications}\label{ss:compact}

There are several modifications that we have to make to the KPV
instanton decay formula due to the fact that we have a compact GKP
geometry rather than a noncompact conifold.

The first and most obvious correction is that gravity  is no longer
decoupled, so we should include the effects of gravitation on the
decay time.  These effects are well known
\cite{Coleman:1980aw,Brown:1987dd,Brown:1988kg}; in appendix
\ref{a:grav}, we work out the specific formula we need.  The decay
time, including gravity (but ignoring the large number of massive
fields in the  compactification), is $t_{decay} \sim \exp[-\Delta S_E]$,
where $\Delta S_E$ is the difference of the Euclidean actions for the
instanton and the initial background state as given in eqn
(\ref{bubbleaction}).  It depends only on the bubble tension, the
initial vacuum energy density, and the change in energy density.
Given two dS states from section \ref{s:parameters}, we just need to
calculate the bubble tension and plug into (\ref{bubbleaction}).

There are also modifications to the tension of the bubble.  The
easiest to calculate is an effect of working in the 4D Einstein frame.
Let us emphasize that we need to work in the 4D Einstein frame to use
the superpotential formalism of section \ref{s:kklt}, and this is also
the frame in which the potential has been calculated.  The Einstein
frame is also the frame used in calculating the instanton decay time.
It is easiest to get this by going to the NS5-brane action
\be\label{SeucNS5} 
S_E= \frac{\mu_5}{g_s^2} \int d^4x \sqrt{\det
g_{\mu\nu}} \delta \int d^3 x\sqrt{g_{S^3}}\ , 
\ee
where $g_{\mu\nu}$
is the 4D pullback of the 10D metric, $\delta$ is the delta
function at the radius of the bubble (with the determinant of the
metric included), and $g_{S^3}$ is the determinant of the metric on the 
$A$ cycle.  The 10D string frame and 4D Einstein frame are
related by $h^{-1/2} g^E_{\mu\nu} = g_s^{-2}e^{6u}g_{\mu\nu}$, so the
NS5 action becomes 
\be\label{NS5tension} 
S_E=2\pi^2 r^3 \tau_5 \ ,\ \
\tau_5 \equiv \mu_5 g_s e^{-9u} \left(\frac{z^{2/3}}{g_s
M}\right)^{3/2}( 2\pi^2) \left( b g_s M \ap\right)^{3/2} =
\frac{b^{3/2}z}{16\pi^3\ap{}^{3/2}g_s^{5/4}\sigma^{9/4}} \ .  
\ee 
(Henceforth $\tau$ is the instanton bubble tension.)  In
the first equality for the tension $\tau_5$, we have separated the
contribution from the conversion to Einstein frame, the warp factor,
and the volume of the $A$ cycle.  We have ignored the contribution to
the action from the NSNS 6-form potential, which KPV showed is
negligible in the thin-wall limit.  Heuristically this is because the
6-form potential only has two legs on the $A$ cycle and the 5-brane
fills the entire cycle, as shown in figure \ref{f:thinwall}.  However, the RR
field strength $F_3$ gives a worldvolume anomaly that requires $M$
D3-branes to attach to the 5-brane.  Here, there are $p$ $\bd$-branes on the
outside and $M-p$ D3-branes on the inside.

The other correction we should make is due to the action for the moduli.
Since the moduli are fixed by the flux superpotential
(\ref{superpotential}), after the NS5-brane bubble changes the flux,
the VEVs of the moduli will change.  Therefore, we need to take into
account the rolling of the moduli to the new vacuum.  We will focus
on the deformation modulus $z$  of the conifold for the following
reasons.  First, it clearly changes  significantly when $K$ changes
(see (\ref{tipwarp})).  Also, for a noncompact conifold, $K$ does not
affect the dilaton or other moduli, so we would  expect that they
would be only minimally affected by a change of $K$ in the compact
case (the other moduli are typically fixed by fluxes on other cycles).
Also, KKLT have shown that the VEV of $\sigma$ does not change much
due to the presence of \bd-branes.  Therefore, since we expect $g_s$
and $\sigma$ to keep roughly the same values before and after the
decay, we expect that they will not roll much, and we will treat them
as constants.  There is actually a significant tree-level potential
for $g_s$ and $\sigma$ when $z$ is not at its VEV,  and we will consider its
effects in the next subsection.  Nevertheless, we expect our estimate of the
contribution from $z$ not to be affected significantly by other moduli.  To
be conservative, one could multiply the contribution from $z$ by a
fudge factor, but we note that we are only making an estimate to begin
with, so we are not quite that careful.

To estimate the tension due to the rolling of $z$, we will assume that 
just inside 
the NS5-brane $z$
is in its original vacuum value outside of the NS5-brane and rolls
quickly to the new VEV inside.  This is probably not the exact
classical solution,  but we will use it and the thin-wall
approximation as an upper limit.  At tree level (where we are
working), we can write the action as \bea S_E(z) &=&
\frac{1}{\kappa_{4}^2}\int d^{4} x \sqrt{g_4} \left[ \mathcal{K}_{z\b
z}\del_\mu z\del^\mu \b z +\k^4 e^{\mathcal{K}}\mathcal{K}^{z\b z}
D_zW \b D_{\b z} \b W\right]\nonumber\\ &=&
\frac{2\pi^2}{\kappa_{4}^2}\int d\xi r^3\left[ \mathcal{K}^{z\b z}
\left( K_{z\b z} \del_\xi z- \k^2  e^{\mathcal{K}/2-i\omega}\b D_{\b
z}\b W\right) \left( K_{z\b z} \del_\xi \b z- \k^2
e^{\mathcal{K}/2+i\omega} D_{z} W\right) \right.\nonumber\\ &&
\left. + \k^2 e^{\mathcal{K}/2+i\omega}\del_\xi z  D_{z} W +\k^2
e^{\mathcal{K}/2-i\omega}\del_\xi \b z \b D_{\b z}\b W\right]\
,\label{SEz} \eea where $\omega$ is some phase (physically, we have to
take it so that the Euclidean action comes out positive because it
started positive definite).   As above, $r$ is the radius of curvature
of the bubble, while $\xi$ is the  radial coordinate corresponding to
proper distance.  This is clearly minimized when only the last two
terms contribute.  Taking the average K\"ahler potential in the
exponential, we get (up to numerical factors of order unity)
\be\label{zcontrib} \tau_z \approx 2\pi^2
e^{\langle\mathcal{K}\rangle/2} \left( |\Delta W| +|\Delta
\mathcal{K}||\langle W\rangle | \right) \ .\ee (This comes from the
definition of the covariant derivative and the chain rule.)  This
derivation is very similar to that of BPS domain walls and is also
used in \cite{Weinberg:1982id}.  Actually, it is easy to  generalize
this estimate to include other moduli, but we will only consider $z$
in the superpotential and K\"ahler potential.  We should note that
$\Delta W$ and $\Delta\mathcal{K}$ are calculated from the inside of
the NS5-brane (where $z$ is not in a vacuum state) to the new vacuum
on the interior of the instanton and not from the original vacuum to
the new vacuum.  Since we are just making an estimate,
$\langle\cdots\rangle$ will be an average value over the region of
variation of $z$.

The change in the superpotential is given entirely by the
superpotential of the conifold just inside the NS5-brane minus $W_0$.
This is because in the vacuum states, the $K$ and $M$ fluxes are (2,1) forms and so do not contribute to the superpotential (see, for
example, \cite{Herzog:2001xk}).  Using the notation and conventions of
\cite{Giddings:2001yu,DeWolfe:2002nn}, we get \bea \Delta W &=& -W(z)
= -\frac{(2\pi)^2 \ap{}^{5/2}}{\kappa_4^8} \left( M
\mathcal{G}(z)-i\frac{K}{g_s} z\right)\nonumber\\ &\approx&
-\frac{(2\pi)^2 \ap{}^{5/2}}{\kappa_4^8} z \left( \frac{M}{2\pi i}\ln
z -i\frac{K}{g_s}\right) \approx  -i\frac{(2\pi)^2
\ap{}^{5/2}}{g_s\kappa_4^8}z\ ,\label{deltaW} \eea where $K$ is the
NSNS flux on the inside of the bubble and $z$ is evaluated outside the
bubble.  This follows from the definitions \be\label{somedefs} \int_A
\Omega = \ap{}^{3/2}z\ ,\ \ \int_B \Omega = \ap{}^{3/2}\mathcal{G}(z)
\ ,\ \ \mathcal{G} = \frac{1}{2\pi i} z\log z
+\mathnormal{holomorphic}\ee and the relation from eqn (\ref{tipwarp})
that $z(\mathnormal{outside}) \approx \exp(-2\pi/g_s M)
z(\mathnormal{inside})$.  To overestimate  $\langle W\rangle$, we will
take  \be\label{avw} |\langle W\rangle| \approx |\wo| +|\Delta W|\ .\ee
 
The K\"ahler potential is significantly more complicated, and, because
we are concerned with a modulus that lives at the bottom of a throat,
we need to take the warp factor into account.  Including warping and
bunching the  K\"ahler potential for all other complex  moduli
together into $\mathcal{K}_c$ (that is, integrals over other cycles),
eqn (\ref{kahlerkklt}) becomes \cite{DeWolfe:2002nn}
\be\label{kahlerwarp} \mathcal{K}(\mathnormal{complex}) =-\log\left[
e^{-\mathcal{K}_c} -\frac{i}{\k^6} \left(\int_A\Omega\int_B \b\Omega h
-\int_A\b\Omega\int_B\Omega h\right) \right]\ .  \ee To compute
$\int_B\Omega h$, we use the trick that the cycles have a monodromy $B
\to B+A$ around $z=0$ (in the same way it was used to find the leading
term in $\mathcal{G}$) and eqn (\ref{tipwarp}) (which is valid at
points both inside and outside the bubble) to expand out
\be\label{kahlerwarp2} \mathcal{K}(\mathnormal{complex}) \approx
\mathcal{K}_c -e^{\mathcal{K}_c} \frac{\ap{}^3 (g_s M)^2}{2\pi\k^6}
|z|^{2/3}\log |z|^2\ .\ee (This is, to our knowledge, the first
calculation of part of a K\"ahler  potential with warping included.)
Actually, there will be other terms in  the $B$ cycle integral, but it
is reasonable to believe that, as in the unwarped case, this is the
leading term that depends on $z$.  Then, using (\ref{kahlerkklt}) and
assuming the complex structure gives small contributions to
$\mathcal{K}_c$,  we get roughly \be\label{kdk}
e^{\langle\mathcal{K}\rangle} \approx \frac{g_s}{16\sigma^3},\ \
\Delta\mathcal{K} \approx -\frac{\ap{}^3 (g_s M)^2}{2\pi\k^6}
|z|^{2/3} \left[ \frac{4\pi}{g_s M} e^{4\pi/3g_s M} +\log|z|^2 \left(
e^{4\pi/3 g_s M} - 1\right)\right]\ .\ee

The total bubble tension is therefore \bea \tau=\tau_5 +\tau_z
&\approx&\frac{b^{3/2}z}{16\pi^3\ap{}^{3/2}g_s^{5/4}\sigma^{9/4}}  
+\frac{2\pi^2 g_s^{1/2}}{4\sigma^{3/2}} 
\left\{ \frac{(2\pi)^2 \ap{}^{5/2}}{g_s\kappa_4^8}z
+\left( \wo + \frac{(2\pi)^2 \ap{}^{5/2}}{g_s\kappa_4^8}z\right)
\right.\nonumber\\ &&\left.\times \frac{\ap{}^3 (g_s M)^2}{2\pi\k^6}
|z|^{2/3} \left| \frac{4\pi}{g_s M} e^{4\pi/3g_s M} +\log|z|^2 \left(
e^{4\pi/3 g_s M} - 1\right)\right|\right\}\ .\label{tension} \eea
With the bubble tension in hand we are now in a position to calculate decay
rates.  However, before moving on we would like to take a closer look 
at subtle issues ignored in the above calculation.  The anxious reader,
fretting over the fate of his or her universe, may skip ahead to
section \ref{s:apply}, and leave the following subsection for
a more careful reading.
 
\subsection{Other Considerations}\label{ss:other}
\subsubsection{D3-brane Migration}\label{sss:d3migrate}

The KPV instanton bubble not only reduces the NSNS flux $K$ and annihilates
\bd -branes, but it also leaves behind D3-branes.  If there are \bd-branes
in other throats, the D3-branes will feel an attraction and roll through the
bulk\footnote{We assume forces due to objects in the bulk, such as D7-branes with gluino condensation \cite{Ganor:1997pe}, can be ignored.  We thank S. Kachru and L. McAllister for discussion on this point.} and into the throat with the \bd-branes.  Eventually, they will annihilate
with the \bd-branes via tachyon condensation.  
If this migration is part of the instanton, then, in many cases, all the 
\bd-branes will be annihilated, leaving a Big Crunch spacetime with negative
energy density.  If there are more \bd-branes to start, the final state could
still be dS.

However, we argue that we should not consider the migration of the D3-branes
to be part of the instanton, but rather as a classical process that occurs
after the bubble nucleates.  Our logic is something like the discussion
of the bounce instanton of quantum mechanics in \cite{Coleman:1978ae};  
the instanton should only tunnel through the barrier to some energy slightly
lower than the initial state, and classical evolution should take over.
Typically, in the thin-wall limit, we just assume that the inside of the
instanton is just the final state.  In our case, though, we expect that
the D3 migration would not be well approximated at all by a thin-wall 
instanton because they are very far from the $\bd$-branes, so the potential is
very flat.  (Contrast this to the case for the $z$ modulus, where the 
gradient of the potential is Planck scale.)  This logic is consistent
with the discussion of Hawking-Moss and related instantons in
\cite{Kachru:2003aw,Banks:2002nm}.

We expect the migration time to be similar to the bubble thickness
for the motion of the D3-branes in the Euclidean description of the 
instanton.  The migration times are larger than the bubble radius for the 
rest of the instanton, so we will treat the D3-brane
migration as a classical process.  In fact, the migration times are 
larger than the initial dS radius itself for the models we consider, 
which is the maximum bubble radius.

In appendix \ref{a:migration}, we estimate the classical migration
time for a single D3-brane migrating from one tip to another.  For the
particular model we examine, $\Delta t_M \sim O(10^{15})$ (in string
units).  As discussed below, decay times for the instantons we are
considering are much larger, $O(\exp[10^9])$.  Thus, inspite of the
fact that total migration will vary a great deal from model to model,
the total decay time 
$\Delta t_{TOT} = \Delta t_{decay} + \Delta t_M \simeq \Delta
t_{decay}$ is relatively unaffected. 

We should note that the classical D3-brane migration followed by 
D3/$\bd$ annihilation could leave a state with negative cosmological
constant.   In cosmology, if the spatial slices have
nonnegative curvature, the FRW constraint equation means that the 
universe cannot actually transition to a negative cosmological constant.
Instead, there is a Big Crunch singularity
\cite{Linde:2001ae,Felder:2002jk,Kallosh:2003mt}.  Though it is  
preferable to end in a dS state after the full decay, this is not
necessary as long as the initial instanton has a lifetime much longer
that the age of the universe.  Note that the instanton, however, ends
in a state of positive cosmological constant, so we avoid the concerns
raised by \cite{Banks:2002nm}.

\subsubsection{Rolling Radius}\label{sss:rollradius}

Now we should go back and examine the classical potential for $\sigma$
that arises because $z$ is away from its VEV.  The behavior of the
radial modulus in flux-generated potentials has been studied in an
attempt to find inflationary behavior in \cite{Frey:2002qc}; we are in
a different regime here because we do not take $z$ to be slowly
rolling.  One point to address is that we cannot actually calculate
the K\"ahler potential  with warping for $z$ excited because it is not
clear if eqn (\ref{tipwarp})  would still hold as $z$ changes.
However, we will assume that it is valid since the starting and
ending points of our evolution are vacuum states for  some values of
the flux $K$.   The key point is that for instantons that go from dS
to dS, the boundary conditions on $\sigma$ mean it should not roll
much, so the following discussion does not apply.  What we are doing
here is comparing instantons with different boundary conditions, one
with $\sigma$ unchanged in the final state and one with
$\sigma\to\infty$ in the final state.

We make the comparison as follows.  The classical potential for
$\sigma$ and $z$ naturally pushes $\sigma$ to large radius  as long as
$z$ is not in its vacuum state (note that this potential is extremely
large compared to the KKLT potential (\ref{nonpertV}),  so we can
ignore the KKLT potential here).   We will make a very rough estimate
of the change in $\sigma$ while $z$ rolls to its vacuum.  If we
believe that $\sigma$ changes enough to get over the barrier of the
KKLT potential before $z$ reaches its vacuum and the classical
potential vanishes, then we  expect dS to Minkowski decays -- mediated
by NS5-branes! -- will dominate over dS to dS decays.  This is
because the classical evolution should have a lower action.
Otherwise, the dS to dS decays will dominate,  at least in the
NS5-brane channel.  We will not say anything else about these dS to
Minkowski decays since they are less computationally tractable and are
somewhat redundant with other decays to large radius.

Now we can roughly estimate the potential for $\sigma$ and $z$.  As
in GKP \cite{Giddings:2001yu}, we work assuming small $z$, which
implies that $\del_z W > (\del_z \mathcal{K}) W$, so we will consider
only the derivative of the superpotential.  As before, the $D_\rho W$ terms
cancel with $-3|W|^2$.  As a final approximation, we take
only the leading terms of the K\"ahler metric for $z$ small.  Thus, we
approximate the potential as \be\label{zpot} V = g_s^4 e^{-12u}
(2\pi)^5\frac{\ap{}^2}{\k^8 (g_s M)^2} \frac{|z|^{4/3}}{|\log |z|^2|}
\left| \frac{M}{2\pi}\log z+\frac{K}{g_s} \right|^2 \ .\ee We have
used  \be\label{kmetric} \mathcal{K}_{z\b z} = -\frac{(g_s
M)^2\ap{}^3}{18\pi\k^6} |z|^{-4/3}\log |z|^2 \ee as the K\"ahler
metric for $z$.  This is singular at $z=0$, but our evolution never takes
$z\to 0$.

To get a very rough estimate of the change in radial modulus $u$
(remember that $\sigma = e^{4u}/g_s$) while
$z$ changes, we approximate that the proper distance in the $u$
direction of moduli space is proportional to the proper distance moved
in the $z$ direction of moduli space.  The proportionality constant is
given by the directional derivative (in the moduli space orthonormal
frame) of the potential.  Using the K\"ahler metric (\ref{kmetric}) to
get the orthonormal frame, we find that \bea \Delta{u} &\approx&
\frac{\Del_{\hat u}V}{\Del_{\hat z}V}  \frac{\sqrt{\mathcal{K}_{z\b
z}} \Delta z}{\sqrt{12}}\label{deltau1}\\ \Delta u &\approx&
\frac{(g_s M)^2\ap{}^3}{18\pi\k^6} |z|^{2/3}  |\log |z|^2| \left(
e^{2\pi/g_s M}-1\right)\label{deltau} \eea up to factors of order
unity.  We have used $\sqrt{\mathcal{K}_{z\b z}}  \Delta z$ for the
proper distance in the $z$ direction.  The factor of $\sqrt{12}$ in
(\ref{deltau1}) comes from the normalization of $u$.

Using the potential graphed in KKLT as a guide, we expect that $\Delta
u$  only needs to be $\gtrsim 0.1$ for the Minkowski decay to
predominate, which is achieved
by $z\gtrsim 10^{-3}$.   As it turns out, we will mainly be interested
in cases with smaller $z$, so we will not consider the 5-brane
mediated dS to Minkowski decays any further.

\subsubsection{Thermal Enhancements}\label{sss:thermal}

Due to the fact that dS has a temperature, we might expect that the
5-branes that make up our instantons should have some nonzero entropy.
Since the exponential of the entropy gives a density of states, the
decay time should be reduced by a factor
$\exp[-\bm{S}(\textnormal{NS5})]$.  This argument was first given in
\cite{Feng:2000if}.  There it was argued that the  brane instantons
probably are out of thermal equilibrium with any matter or radiation
in the cosmology, so they should have a temperature corresponding to
the dS temperature.  However, whether the temperature should be the
initial dS temperature, final dS temperature, or the geometric mean
was undetermined.  It is now clear \cite{Fabinger:2003gp} 
that the brane would be
in equilibrium with the initial dS because it corresponds to
accelerating observers in the two dS spacetimes.\footnote{We thank the
authors of \cite{Fabinger:2003gp} for sharing their results with us prior to
publication.}

We will, however, neglect this effect.  The bubble temperature
is just the inverse radius, 
$T=1/(2\pi r)$ \cite{Fabinger:2003gp}.
Therefore, the temperature is not high enough to excite the ``Kaluza-Klein''
modes of the bubble much, and the entropy would access only the zero-mode
quantum mechanics.  We expect that the enhancement factor would be
relatively weak, therefore.

\section{Calculation of decay times}\label{s:apply}

Throughout this paper, we have mainly discussed the KPV instantons as
CDL thin-wall instantons.  However, they contain an NS5-brane, which
makes them also of the membrane class of instantons studied by
\cite{Brown:1987dd,Brown:1988kg}.  In appendix \ref{a:grav}, we demonstrate
the equivalence of these two formalisms by showing that they give the same
decay rate given initial and final cosmological constants and instanton
tension.  

Using the results of section \ref{s:kpvinst} and appendix \ref{a:grav},
we are able to calculate decay rates.  For illustrative purposes let's
first consider a model with a single KS throat.  In particular, for 3
\bd-branes sitting at the tip of a throat with 
$K = 12,\ M=87$, one finds that the probability per unit volume for
NS5-brane mediated decay is $P \sim \exp (-10^{19})$.  Decays to
decompactification are much faster, $P \sim \exp (-10^{17})$.  We
expect this to generally be the case for single throat models.
Moreover, as discussed in \cite{Banks:2002nm}, since all single throat
decays will have $\Lambda < 0$ in the final state, the instantons
mediating these decays might not exist.  It is for this reason we have
chosen to focus on models with 2 KS throats, which, after the initial
decay, have $\Lambda >0$.  

What follows is a discussion of the decay rates for several different
two throat models.  Table \ref{t:models}
shows the the fluxes and number of $\bd$-branes, $p_i$, in each throat. 
In each model the initial KPV instanton occurs in throat 1. This
decay is driven by the notably small value of $z_1$, which makes the tension
very small.  Note that we have specifically chosen models where this
is the case. The change
in  and resulting value of  the
cosmological constant ($\Delta \Lambda$ and $\Lambda_-$ respectively), 
due to KPV decay, are also given in table 
\ref{t:models}.  The small value of $z_1$ corresponds to small $|\Delta\Lambda|$, 
which would increase the decay time, but this effect is compensated by 
the small bubble tension. How the decay rate depends on these values is
given explicitly by eqn (\ref{bubbleaction}).  We list the tensions and decay
times\footnote{Note
that these are the decay times for a unit volume, i.e. 
$t_{decay} = P^{-1}$} for the KPV instantons in table \ref{t:tensions}, along with tensions and decay
times for two other decay modes discussed in section \ref{s:compare}
below. Note that the lifetimes for these models are $\sim \exp
(10^{9})$, where as the age of the universe (times the horizon
volume) is $\sim \exp(10^3)$, so even the most anxious reader can now
relax and enjoy the rest of the paper.

\begin{table}[t]
\begin{center}
\begin{tabular}{|c|c|cccc|cc|cc|}
\hline 
Model & $p_1,p_2$ & $K_1$  &  $M_1$  &  $K_2$  &  $M_2$  
 &  $z_1 \times 10^{17}$  & $z_2 \times 10^{5}$  
 &$\Delta \Lambda \times 10^{31}$ &  $\Lambda_-  \times 10^{17}$  \\
\hline
\hline  
1 & 1,1 & 9  & 15 & 3 & 19 & 4.2 & 4.9 & 3.9 & 69 \\
2 & 1,1 & 9  & 15 & 4 & 26 & 4.2 & 6.3 & 3.9 & 2.7 \\
3 & 1,1 & 9  & 15 & 9 & 69 & 4.2 & 28  & 3.9 & 4.5 \\
4 & 1,5 & 9  & 15 & 8 & 51 & 4.2 & 5.2 & 3.9 & 4.5  \\
5 & 1,5 & 9  & 15 & 13 & 91 & 4.2 & 13 & 3.9 & 7.6 \\
\hline
\end{tabular} 
\end{center}
\caption{\label{t:models}Models and Cosmological Constants}
\end{table}

\begin{table}[t]
\begin{center}
\begin{tabular}{|c|cc|ccc|ccc|}
\hline
Model & ${(\tau/\tau_c)}\ $ & $\ln (t^{KPV}_{decay}) \times 10^{-9}$
      & KKLT:& ${(\tau/\tau_c)}\ $  
      & $\ln (t^{KKLT}_{decay}) \times 10^{-18}$
      & T2T:& ${(\tau/\tau_c)}\ $  
      & $\ln (t^{T2T}_{decay}) \times 10^{-18}$ \\ 
\hline
\hline  
1 & 0.163 & 0.66 & & 1.8 & 0.32 & & 24970 & 0.35 \\
2 & 0.164 & 86   & & 7.7 & 8.9  & & 24257 & 8.9  \\
3 & 0.164 & 40   & & 5.9 & 5.2  & & 16512 & 5.2  \\
4 & 0.163 & 3.7  & & 2.9 & 1.1 & & 34054 & 1.1 \\
5 & 0.164 & 18   & & 4.6 & 3.1  & & 33517 & 3.1  \\
\hline
\end{tabular}
\end{center}
\caption{\label{t:tensions}Tensions and Decay Times}
\end{table}

For each of the models discussed above, although the initial instanton
decay yields a spacetime with positive cosmological constant, the ensuing
D3-brane migration results in a negative cosmological constant, a situation
which, as discussed in \cite{Linde:2001ae,Felder:2002jk,Kallosh:2003mt}, 
ultimately leads to a Big
Crunch singularity.  Note, as previously mentioned, this is a
classical process and thus avoids arguments given against
instanton decays to negative $\Lambda$ \cite{Banks:2002nm}.
The total migration time, as shown in appendix \ref{a:migration}, is
negligible compared to the decay time\footnote{Note, however, that it
is long compared to the string scale, $\mathcal{O}(10^{15})$.} and
will thus be ignored.  We should also note that, although it seems difficult
to find two throat models with a positive cosmological constant after
D3/\bd\ annihilation, it should be possible to construct 
multiple ($>2$) throat models that end in dS.

\section{Comparison to other decay modes}\label{s:compare}
The KPV instanton is just one of several avenues by which these dS
vacua can decay.   
One particular mode, thoroughly studied in 
\cite{Kachru:2003aw, Giddings:2003zw} and reviewed at the end of
section \ref{s:kklt}, is tunneling 
to decompactification (in the CDL formalism).  
In these decays, or for any decay
in which $\Lambda_- = 0,\ \Delta S_E$ takes a particularly
simple form,
\be\label{Sdcmpt}
\Delta S_E = -\frac{S_0}{\left(1+\tau_c^2/\tau^2\right)^2} \ .
\ee
For comparison purposes, we have calculated the CDL tensions and decay times
$(t^{KKLT}_{decay})$ for five models discussed above.  These are also
listed in table \ref{t:tensions}.
Note that in each model the tensions are super-critical, 
$\tau/\tau_c > 1$.  This will in fact always be true for decays to 
decompactification since,
\be\label{supercrit}
\frac{\tau}{\tau_c} = 
\frac{1}{\sqrt{4  V(\phi_+) /3}} \int_{\phi_+}^\infty d\phi \sqrt{2 V(\phi)} \
\geq \  1 \ ,
\ee
for any $V(\phi)$ whose barrier width (in
string/Planck units for our normalization) is greater that $\sqrt{2/3}$.  
Noting that $S_0 < 0$, it is clear from (\ref{Sdcmpt}) 
that the lifetime, $t_{decay} \sim \exp(-\Delta S_E)$, increases
with $\tau/\tau_c$.  Though the story is 
more complicated when comparing to decays with $\Lambda_- \neq 0$, this
will generally still be the case, and it is this fact which drives 
$t^{KKLT}_{decay}$ to be much greater that $t^{KPV}_{decay}$.  
Take careful note that table
\ref{t:tensions} lists the logs of the decay times.  For these
models $t^{KKLT}_{decay}/t^{KPV}_{decay} \sim
\exp{(10^8)}$!  These KPV instantons are, in technical terms, much much much
faster.  It is possible to find super-critical KPV instantons in
which $t^{KKLT}_{decay} < t^{KPV}_{decay}$.  However, these require larger
$z$ in the decaying throat, leading to larger initial cosmological constant
and slower decay times. 

Another particularly simple decay mode occurs in models with multiple 
KS throats.  The potential energy of a \bd-brane 
is proportional to $h^{-1}(r)$, the inverse warp factor given by eqn (\ref{tipwarp}),
which is locally minimized at the tip of each throat.  However, the 
energy is lower still at the tip of other throats with smaller $z$.  \bd-branes can therefore tunnel from one throat to another.  On the other hand, $h \sim 1$ in the bulk, 
presenting a substantial potential barrier through which to tunnel. These instantons are similar to the glueball decays considered in \cite{Dimopoulos:2001qd,Dimopoulos:2001ui}.

As in previous examples, we consider models with two KS throats.  
The \bd-brane portion of the total potential is initially 
(cf. eqn (\ref{MdeltaV}))
\be
\label{t2tdeltaV}
\delta V = \frac{2 \mu_3}{\sigma^3} h^{-1}(\rt_1) 
\ p_1 + \frac{2 \mu_3}{\sigma^3} h^{-1}(\rt_2) \ p_2 \ .
\ee 
After the tunneling occurs, the form of $\delta V$ is unchanged except for
$p_1 \to p_1 + 1$ and $p_2 \to p_2 - 1$.  These decays have little
effect on $\sigma$, and thus $\sigma$ will be treated as a 
constant throughout 
this calculation.  

To find the decay rate, we compute the instanton tension from the
Euclidean brane action in the thin-wall limit using \cite{Coleman:1980aw}:
\be\label{CDLtension}
\tau = (2\pi\sqrt{\sigma})^{3/2}g_s^{1/4}\ap\int_{\rt_1}^{\rt_2} dr \ 2
\sqrt{\frac{\mu_3}{\sigma^3} [h^{-1}(r)-h^{-1}(\rt_1)]} \ .  
\ee 
The prefactor is from the conversion between $r$ and a canonically 
normalized scalar in the 4D Einstein frame.
Note that here we are using rescaled coordinates so $e^{2u}$ does not
appear in the metric (\ref{conifold}).
We then plug $\tau$ and $\Lambda_\pm = \kappa_4^2 (V + \delta V_\pm)$ into 
eqn (\ref{bubbleaction}) to obtain the ``throat-to-throat'' decay
time $(t_{decay}^{T2T})$.  Once again, the these decay times and
tensions are listed in table \ref{t:tensions}.  Note, however, that these
instantons tunnel to negative cosmological constant; while they would
be ruled out by \cite{Banks:2002nm}, they are not forbidden by 
the original calculation of \cite{Coleman:1980aw}.
As with the KKLT decays, the tensions are uniformly
supercritical, and $t_{decay}^{T2T}$ is remarkably similar to 
$t_{decay}^{KKLT}$.  Indeed, we expect these decays to be super-critical
because the conifold throats are long in string units, giving a wide
potential barrier.  Moreover, taking the
limit $\kappa_4^2 \tau \gg \Lambda_+$ 
in eqn (\ref{bubbleaction}), it is easy to see that,
\be
(\Delta S_E)_{T2T} \approx \frac{24 \pi^2}{\Lambda_+} \ ,
\ee    
and thus from (\ref{backgroundaction}) and (\ref{Sdcmpt}), one can see
that $t_{decay}^{KKLT} \sim t_{decay}^{T2T}$.

The reader should remember that we are working only in the thin-wall
limit and that Hawking-Moss instantons can also give significant 
contributions to the decay rate.  However, in the models we have described,
the dS to dS decays are subcritical, so the Hawking-Moss contributions
seem unlikely to change our qualitative results; KKLT found
that Hawking-Moss instantons begin to dominate over thin-wall instantons only when
$\tau \sim \sqrt{V(\varphi_1)} > \tau_c$.
  
The KPV instanton deals only with changes to fluxes and branes in 
eqn (\ref{tadpole}).  One might
speculate about processes which could involve changes to $\chi(X)$, 
or the induced D3 charge on
wrapped $(p,q)$ 7-branes in the IIB.  From the F-theory viewpoint this
would obviously involve topology change.  While one could consider
non-trivial D7-brane worldvolume gauge fields undergoing a small instanton
transition and emitting D3-branes into the bulk, we know of no analog for
non-trivial curvature on a four-cycle.  A possible $\chi$ changing instanton
would involve D7-branes unwrapping a particular 4-cycle and wrapping a 
different one; however, these 4-cycles would be homologous unless the 
D7-brane can tear, so the induced D3-brane charge would remain the same.
However, it may be
interesting to explore whether $\chi$-changing mechanisms are possible.

\section{The End of the World (And This Paper) As We Know It}
\label{s:conclusion}

The decays considered in this paper in a very real sense would represent
the end of the universe for anyone unfortunate enough to experience 
one.\footnote{Apologies to REM for the section title.}
Note, however, unlike the decays in CDL, even when D3/\bd\ annihilation 
following a KPV decay results in a Big Crunch, lifeforms might be
capable of knowing joy for $10^{-28}$ seconds while the D3-branes migrate across the compact manifold.  We can all take comfort in the fact that even the fastest decays we 
consider have decay times incredibly greater than the age of our universe.
Assuming that our calculations hold even approximately for a compactification
with a realistic cosmological constant, we will have to worry about the
death of the sun long before the death of the universe.

Of interest, however, is the fact that we constructed decay modes other 
than the straightforward decay to decompactification discussed in
\cite{Kachru:2003aw,Susskind:2003kw,Giddings:2003zw}.  In fact, we found
it easy to construct NS5-brane mediated decays that occur much more
rapidly than the decompactification decays.  
We reiterate that the NS5-brane decays can have a subcritical tension.
It is also noteworthy that
the final state of many decays is not 10D Minkowski spacetime is instead dS 
or a space with negative cosmological constant which ends in a Big Crunch.  
In fact, depending on the region of parameter space, we found that 
decays mediated by NS5-branes can end in dS, 10D Minkowski, or with
negative cosmological constant, without considering other decay channels!
The lesson is that, even in the KKLT models, there are many different 
metastable vacua and many different possible decay modes.

\begin{acknowledgments}
We would like to acknowledge comments from
H. Ooguri and M. Schulz and very helpful discussions 
with O. deWolfe, B. Freivogel, S. Giddings, T. Hertog, C. Herzog,
G. Horowitz, S. Kachru, L. McAllister, J. Polchinski,  and E. Silverstein.   
Again,
we thank M. Fabinger and E. Silverstein \cite{Fabinger:2003gp} 
for sharing their results with us prior to publication.  
The work of A.F.\ was supported by National Science Foundation grant
PHY00-98395.  The work of M.L.\ was supported by Department of Energy 
contract DE-FG-03-91ER40618.  The work of B.W.\ was supported by National 
Science Foundation grant PHY00-70895.
\end{acknowledgments}

%
%
\appendix

\section{Migration of D3-branes}\label{a:migration}
The KPV decay leaves D3-branes at the tip of the KS throat where it occurs.
In this appendix we analyze their subsequent classical motion in 
configurations with multiple KS throats.  The D3-branes, produced by a 
decay in one throat (Throat 1), are attracted by \bd-branes in another 
throat (Throat 2), migrate across the compact manifold $M$,
and eventually annihilate the $\bd$-branes.  
Here we work in the SUGRA limit to approximate
the total migration time, in a two-throat geometry.  As discussed in section
\ref{s:apply}, the KPV/CDL decay times are so large that the
migration times have little effect.  This appendix, therefore
serves largely to show that one may, in fact, ignore the migration time
and to illuminate how the migration itself proceeds.

It will be assumed that the back-reaction of the migrating D3-branes 
is negligible, the proper velocity of the branes remains small, and that the
majority of the travel time comes from the two throats (i.e. the time
through the bulk of the CY may be ignored).

We turn our attention once more to the metric (\ref{conifold})
(with the overall scale of the manifold scaled back in).  In 
\cite{Klebanov:2000hb} it was shown that the $F_3$ flux wrapped on the A-cycle
smoothly deforms the tip of the conifold.  
Though we will find that the majority of the travel time comes from
the tip of throat 1, we may ignore the most of the details coming
from the deformation of the conifold since the motion is
assumed to be radial.  We will use the undeformed metric and, when
working near the tip, multiply the warp factor $h$ by an
overall constant $\sim 0.4$ to account for the deformation\footnote{One 
finds this correction by comparing  
the ``near tip'' warp factor found in \cite{Klebanov:2000hb} to 
the naive limit of the undeformed Klebanov-Tseytlin geometry.}.  
This has little effect on the final result, however it was such a
trivial correction it seemed silly not to include it.  
The warp factor, away from the tip, is
\bea\label{warpfactor}
h    &=& \frac{L^4}{r^4} \ln(r/r_s) \\
r_s  &=& r_0 \ \exp (-\frac{2 \pi(N+p)}{3 g_s M^2} - \frac{1}{4} ) \\
p  &\equiv& \# \mbox{ of  \bd-branes} \ ; \ \  r_0^2 = 3/2^{5/3} \ \ . 
\eea
Due to the deformation of the conifold discussed above we will only be
interested in the region, $\rt = r_s \exp (1/4) \leq r \leq r_0$.
Note that this avoids the naked singularity at $r = r_s$.   

The action for the D3-branes is
\be
S_3 = -\frac{\mu_3}{\gs} \int d^{4}\xi \ \sqrt{-\gt} \ + 
\mu_3 \int_{\Sigma(D3)} \cf
 \ .
\label{DBI}
\ee
The Ramond-Ramond potential $\cf$ depends only on the
radial distance $r$:
\be 
\cf = \frac{f(r)}{g_s}  dt \wedge dx^1 \wedge dx^2 \wedge dx^3 
\ee
where $t \equiv x^0$.  Working in the
gauge $\xi^0 = \tau(t)$ and $\xi^i = x^i$, the Lagrangian becomes,
\be\label{fullL}
\mu_3^{-1} \fL  = -  \frac{\hi}{\gs}(\sqrt{\tmhr}) + f(r) \td .
\ee
Assuming that the proper velocity is small,
\be\label{Lag}
\mu_3^{-1} \fL  \simeq \frac{1}{2} \frac{\rds \td^{-1}}{g_s}
                       - \frac{(h^{-1} - g_s f(r))}{g_s} \td 
\ee
It is easy to check that this a valid approximation. In particular, one
only needs to consider throat 2, since this is where the
D3-branes are moving fastest.  One can show that 
$p << (g_s M^2)/8$ will insure that (\ref{Lag}) is valid.
   
Since $t(\tau)$ is a cyclic variable, we know that 
$\partial \fL / \partial \td \equiv - E/g_s $ is constant, leaving us with
\be\label{Eeqn}
E = \frac{1}{2} (\frac{\partial r}{\partial t})^2 + V(r)\ ; \ \
V(r) \equiv (h^{-1} - f(r)) \ \ ,
\ee
and the travel time through a single throat is therefore,
\be\label{traveltime}
\Delta t = \pm \int_{r_0}^{\rt} \frac{dr}{\sqrt{2 (E - V(r))}} \ \ .
\ee
The $+/-$ corresponds to branes traveling into/out of the throat.

Note that the time here is a coordinate time, but we will see that it is
so small compared to decay times that we do not need to worry about
conversion to proper time in the 4D Einstein frame.

\subsection{Throat 1}
The D3-branes, produced at rest in throat 1, feel a slight gravitational
attraction to the \bd-branes at the bottom of throat 2.  Since the
gravitational attraction is weakest while in throat 1, one suspects
the majority of the migration time it comes from throat 1.  We show
below that, in fact, throat 2 can be ignored completely.  This also
justifies not including the travel time through the bulk of the CY.

The Ramond-Ramond potential, $\cf = h_1^{-1}/g_s$, 
is not effected by charge contained in throat 2.  Physically, 
this is due to the charge being screened; mathematically, we are
working on a compact manifold and may consider just the charge
enclosed in throat 1.  However, the geometry does, albeit slightly,
know about what is happening in throat 2.

In order to get an estimate of the migration time we will make the 
(perhaps bold) assumption that, as with most multi-pole solutions in
gravity, the warp factor is changed by an additive factor,
\be
h_1(r) \longrightarrow h_1(r) + \delta(r); \ \ 
\delta(r) \equiv - \frac{27 \pi \ap}{(2 r_0 -r)^4} 
\bigl(N_2 + p + 3 g_s M^2 / 8 \pi \bigr) \ .
\ee
Here, the subscripts indicate the throat in which the fluxes are contained, 
$p$ is the number of $\bd$-branes in throat 2 (there are no $\bd$-branes
in throat 1).  Expanding to first order in $\delta$, we see that 
\be\label{pot1}
V(r) = - \frac{27 \pi \ap}{(2 r_0 -r)^4} 
\bigl(N_2 + p + 3 g_s M^2 / 8 \pi \bigr) h_1^{-2}(r) \ \ .
\ee 
%
%
\begin{figure}
\begin{center}
\includegraphics{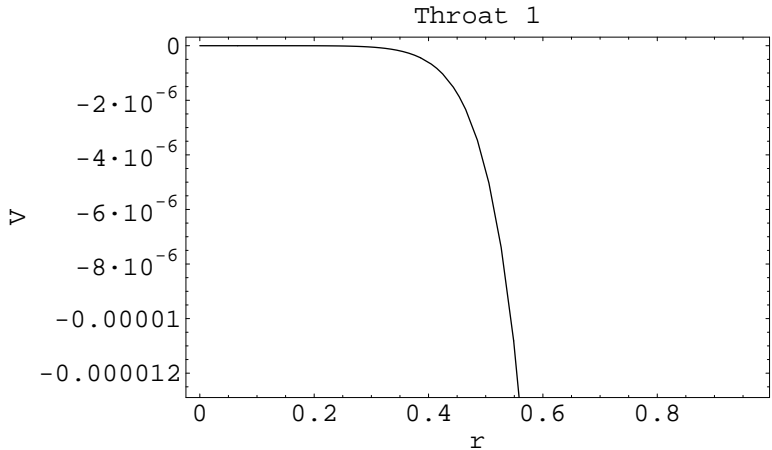}
\end{center}
\caption{Potential in throat 1 (Model 1)} 
\label{f:v1plot}
\end{figure} 
Figure \ref{f:v1plot} shows the this potential.  Note that the
majority of the time will be spent at the tip of this throat.  We have
been unable to evaluate the integral (\ref{traveltime}) explicitly.
Numerical methods also proved difficult, due to the singular behavior
of the integrand as $r\rightarrow \rt$.  This results
from the fact that the D3-branes are produced at rest.  However, one may
gain control of the situation by linearizing the potential near
$r = \rt$ and integrating away from the problematic singular point.
It is in this limit that we multiply $h_1$ by the numerical
factor $\sim 0.4$ discussed above.
Once a safe distance away from $r=\rt$, which for technical reasons
coming from the linearization of (\ref{pot1}) we take to be 
$r \sim \rt + z r_0$, one may
evaluate the rest of the integral (\ref{traveltime}) numerically.
For model 1, one finds that
\be\label{time1}
\Delta t_1 = 1.5 \times 10^{16} \ .
\ee

\subsection{Throat 2}
For completeness, we will determine time spent in throat 2,
subsequently showing that it is of no importance.
In order to deduce $\cf$, recall that $\int \star d \cf \sim
(N_{eff}-p)$.  Plugging in the appropriate constants, this leaves us
with
\be
r^5 h^2 \partial_r f = (27 \pi {\ap}^2) (N_{eff} - p) \ \ .
\ee
Using what we know from the KS geometry, we define $f = h^{-1} + V$,
where the potential, $V(r)$, satisfies
\be\label{Eeqn2}
\partial_r V = \frac{(27 \pi {\ap}^2 g_s) p}{L^5} 
\frac{r^3/L^3}{(\ln(r/r_{s+}))} \ \ .
\ee 
This can be integrated and gives a solution depending on exponential-
integral functions. 
%
%
\begin{figure}
\begin{center}
\includegraphics{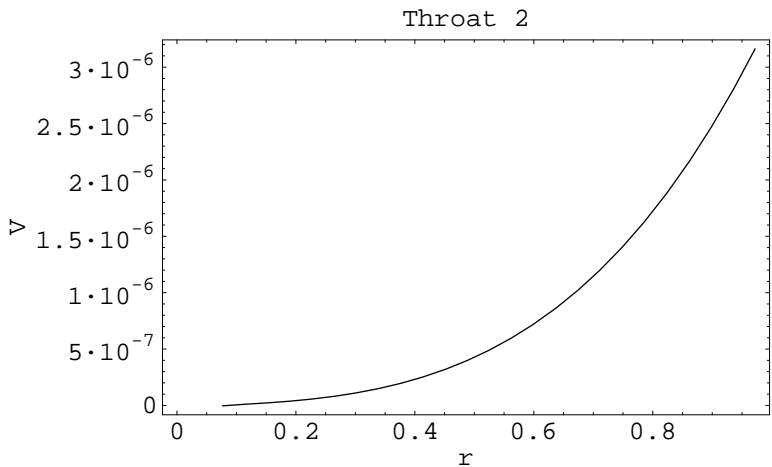}
\end{center}
\caption{Potential in throat 2 (Model 1)} 
\label{f:v2plot}
\end{figure} 
The potential is shown in figure \ref{f:v2plot}.  Numerically
integrating (\ref{traveltime}) for throat 2 gives
\be\label{time2}
\Delta t_2 \approx 9.9 \  ,
\ee
which is clearly negligible compared to (\ref{time1}).

\section{Inclusion of Gravity in Bubble Nucleation}\label{a:grav}

In recent literature, there has been some confusion concerning the 
relation of two formalisms for studying thin-wall instantons.  The method
of Coleman and De Luccia (CDL) \cite{Coleman:1980aw,Banks:2002nm} 
describes smooth
instantons in the limit of large radius of curvature; this formalism was
used by \cite{Kachru:2003aw} to argue that bubbles will always nucleate in a dS background before the recurrence time.  Alternately,
however, we could imagine that the bubble wall is truly an infinitesimally
thin membrane, such as a D-brane, with a delta function stress tensor.  
This type of configuration was studied by Brown and Teitelboim (BT)
\cite{Brown:1987dd,Brown:1988kg}.  In a description of dS solutions in 
noncritical string theory \cite{Silverstein:2001xn}, \cite{Maloney:2002rr} uses
the BT formalism to argue that there is a critical tension above which the
bubble occupies more than half the original de Sitter sphere and above which the 
decay time changes behavior as a function of the bubble tension.  Additionally,
\cite{Maloney:2002rr} claims that the decay time is of order the recurrence time at the critical tension.  

In this appendix, we show that the CDL and BT formalisms actually agree; 
this is reassuring, since even D-branes should
be described as smooth objects in a complete version of string theory.
Our results show that the decay time is always less than the recurrence time, as in 
\cite{Goheer:2002vf,Kachru:2003aw,Giddings:2003zw,Susskind:2003kw}.  Additionally, we confirm that more than half the original de Sitter sphere decays above the critical tension, but
we show that (due to some technical considerations) the decay time is 
actually a smooth function of the bubble tension.  We work in a 4D effective
theory throughout.

We begin by describing the two formalisms.  
In both CDL and BT, the nucleation/decay time is
given by exponentiating the difference of the (Euclidean) bubble and 
background actions.  Therefore, the exterior of the bubble, which is 
approximated by the background, contributes nothing in both formalisms.

At that point, CDL note that, since both bubble and background have the same
behavior at infinity, they can integrate some terms in the Ricci tensor
by parts.  After determining the bubble tension $\tau$ as a functional of
the potential, they find that the action is
\be\label{cdlaction}
\Delta S_E = 2\pi^2 r^3 \tau  +\frac{12\pi^2}{\kappa_4^2}\left\{
\frac{1}{\Lambda_-} \left[ \left( 1-\frac{\Lambda_-}{3}r^2\right)^{3/2}-1
\right]-\frac{1}{\Lambda_+}\left[ \left( 1-\frac{\Lambda_+}{3}r^2
\right)^{3/2}-1\right]\right\}
\ee
where $\Lambda_\pm = \kappa_4^2 V_\pm$ are the potential outside and inside
the bubble respectively (this is a combination of eqns (3.11) and (3.13) from \cite{Coleman:1980aw}).  Minimizing this action with respect to the bubble
curvature radius $r$ gives the decay rate.

On the other hand, BT cannot use the same integration by parts because the
infinite stress of the bubble wall separates the interior and exterior 
regions.  Instead, the bubble action must include extrinsic curvature
terms; it is these terms that will explain the apparent contradiction between
CDL and BT formalisms.  The extrinsic curvatures of the interior and
exterior regions are
\be\label{curvature}
K_\pm = -3\sigma_\pm \left( \frac{1}{r^2} -\frac{\Lambda_\pm}{3}\right)^{1/2}
\ee
where $\sigma_\pm=1$ if the radius of curvature of the outside/inside
region of the bubble is increasing toward the exterior of the bubble and 
is $-1$ if the radius is decreasing.  Since the Ricci scalar in the bubble is
given by the cosmological constant, the action just becomes
\be
\Delta S_E = 2\pi^2 \tau r^3 + \frac{2\pi^2}{\k^2} r^3\left( K_- -K_+
\right) -\frac{1}{\k^2} \left( \Lambda_- \mathcal{V}_- -
\Lambda_+ \mathcal{V}_+\right)
\ . \label{btaction}
\ee
The interior volumes for the bubble and background are given by (for either
sign of the cosmological constant)
\be\label{volumes}
\mathcal{V}_\pm = 2\pi^2 \left(\frac{3}{\Lambda_\pm}\right)^2 \left\{
\frac{1}{3}\left[\sigma_\pm \left( 1-\frac{\Lambda_\pm}{3}r^2\right)^{3/2}
-1\right] -\left[ \sigma_\pm \left( 1-\frac{\Lambda_\pm}{3}r^2\right)^{1/2}
-1\right]\right\}  \ .
\ee
It is algebraically simple to see that the extrinsic
curvature terms combine with the square root terms from the volume to give
exactly the CDL action (\ref{cdlaction}) up to the signs $\sigma_\pm$.
The reason \cite{Maloney:2002rr} found a different
action is that they omitted the extrinsic curvature terms.

Now we should see why the CDL result should actually have the 
signs $\sigma_\pm$.
For a de Sitter background, the terms in square brackets of eqn 
(\ref{cdlaction}) come from integrals
\be\label{integrals}
\int_0^{\xi (r)} d\xi\ r\left(1-\frac{\Lambda_\pm}{3} r^2\right)\ ,
\ee
which CDL evaluate by replacing $d\xi = dr (1-r^2 \Lambda_\pm/3)^{-1/2}$.
However, as they note, $r=\sqrt{3/\Lambda_\pm} \sin [\sqrt{\Lambda_\pm/3}\xi]$,
so $\xi (r)$ is double-valued.  In fact, the correct integral is
\be\label{newint}
\frac{3}{\Lambda_\pm} \int_{\sigma_\pm(1-r^2 \Lambda_\pm/3)^{1/2}}^1 dy\ y^2
\ee
which just introduces a factor of $\sigma_\pm$ in the $(\cdots)^{3/2}$ terms.
This precisely agrees with the BT results.  In this paper, we will be 
concerned only with the case $\sigma_- = 1$, and $\sigma_+ = -1$ only for 
$\Lambda_+ >0$ and tension above critical.

To find the radius of the bubble given the two cosmological constants and
the tension, we could minimize the action with respect to $r$.  However,
it is easier to use the Israel matching condition across the bubble wall, 
which has trace $K_+ - K_- = (3/2)\k^2\tau$.  The answer is given by 
\cite{Maloney:2002rr} and can be written as
\be\label{radius}
\frac{1}{r^2} = \left(\frac{\k^2\tau}{4}\right)^2 +\frac{\b\Lambda}{3}
+\left(\frac{\Delta\Lambda}{3\k^2\tau}\right)^2\ ,\ \ 
\b\Lambda= \frac{\Lambda_+ +\Lambda_-}{2}\ ,\ \Delta\Lambda = \Lambda_- -
\Lambda_+\ .\ee
It is tedious but straightforward to check that this matches the result
from minimizing the action.
The maximum radius occurs at critical tension
\be\label{critical}
\k^2\tau_c = \left( \frac{4}{3} |\Delta\Lambda|\right)^{1/2}\ee
and is $1/r^2 = \Lambda_+/3 = 1/R^2_{\textnormal{\scriptsize dS}}$ for positive
initial cosmological constant. Note that for vanishing initial vacuum energy,
gravity stabilizes the false vacuum for tension bigger than
critical, as in \cite{Coleman:1980aw}.  Also, for negative initial 
cosmological constant, the radius becomes infinite for tensions lower than
critical.  We will concern ourselves only with initial de Sitter spacetimes,
so we do not face some of the concerns raised by \cite{Banks:2002nm} about
decays of Minkowski and AdS spacetimes.  

\begin{figure}[t]
\begin{centering}
\subfigure[Lines, from top to bottom, are at 
$dL =-0.1,-0.3,-0.55,-0.75,-1,-1.5$.]{
\includegraphics[scale=0.75]{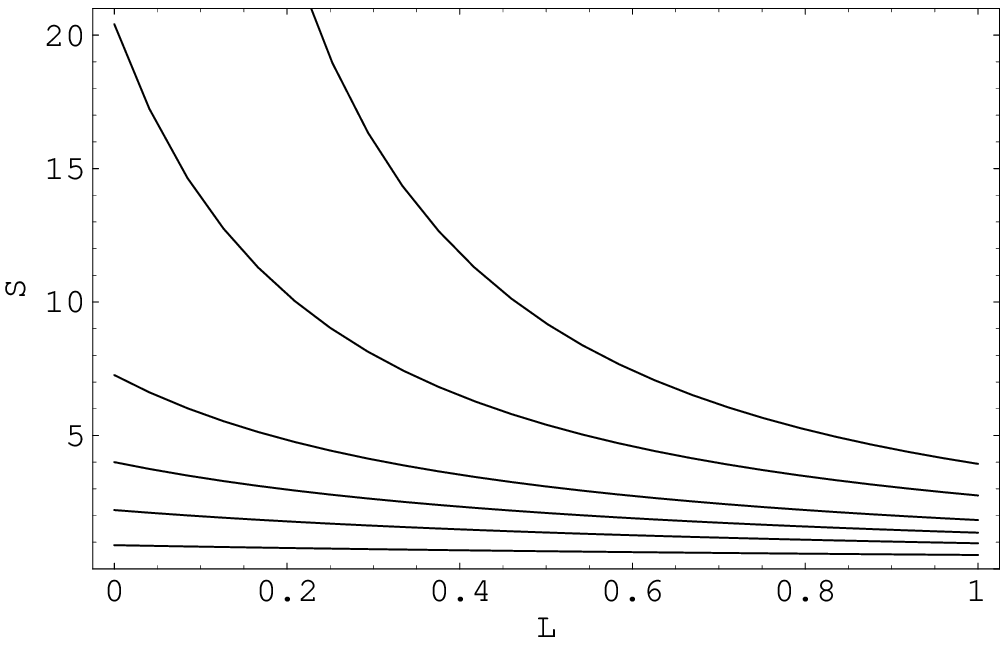}}
\subfigure[Dependence on both variables.]{\includegraphics[scale=0.75]
{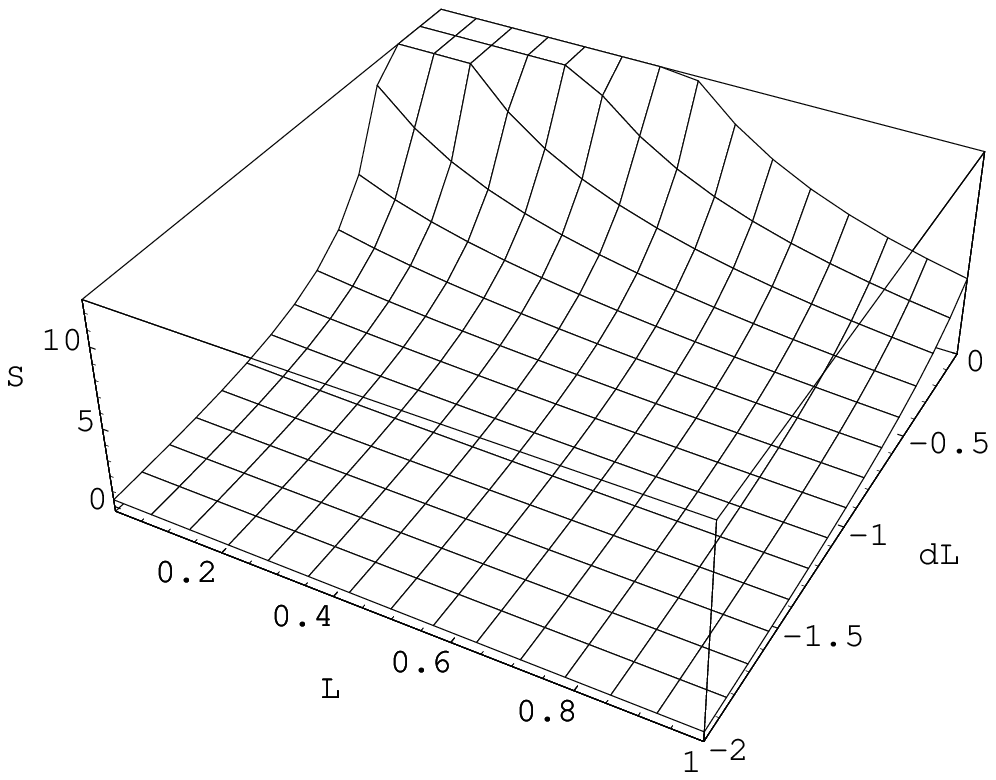}}
\caption{\label{f:cdlbt}The bubble minus background action as a function of
the cosmological constants.  The variables are $S=\k^4\tau^2 \Delta S$,
$L=\Lambda_-/\k^4\tau^2$, and $dL=\Delta\Lambda/\k^4\tau^2$.}
\end{centering}
\end{figure}

We will finally write down the action for the bubbles:
\bea
\Delta S_E &=& 2\pi^2 r^3\left\{ \tau +\frac{6}{\k^2\Lambda_+ \Lambda_-}
\left[ \frac{\Delta\Lambda}{r^3} 
+\Lambda_+ \left(\left(\frac{\k^2\tau}{4}\right)^2 
-\frac{\Delta\Lambda}{6}
+\left(\frac{\Delta\Lambda}{3\k^2\tau}\right)^2\right)^{3/2}
\right.\right.\nonumber\\
&&\left.\left. -\sigma_+\Lambda_- \left(\left(\frac{\k^2\tau}{4}\right)^2 
+\frac{\Delta\Lambda}{6}
+\left(\frac{\Delta\Lambda}{3\k^2\tau}\right)^2\right)^{3/2}
\right]\right\}\label{bubble1}\\
&=& \frac{2\pi^2}{\left(\left(\frac{\k^2\tau}{4}\right)^2 +\frac{\Lambda_- 
-\Delta\Lambda/2}{3}
+\left(\frac{\Delta\Lambda}{3\k^2\tau}\right)^2\right)^{3/2}}
\left\{ \tau +\frac{6}{\k^2\left(\left(\Lambda_- -\frac{\Delta\Lambda}{2}
\right)^2
-\frac{\Delta\Lambda^2}{4}\right)}\right.\nonumber\\
&&\left.\times
\left[ \Delta\Lambda \left(\left(\frac{\k^2\tau}{4}\right)^2 +
\frac{\Lambda_- -\frac{\Delta \Lambda}{2}}{3}
+\left(\frac{\Delta \Lambda}{3\k^2\tau}\right)^2\right)^{3/2}
\right.\right.\nonumber\\
&&\left.\left. +\left(\Lambda_- - \Delta\Lambda\right)
\left(\frac{\k^2\tau}{4}-\frac{\Delta\Lambda}{3\k^2\tau}\right)^3
+\Lambda_-
\left(\frac{\k^2\tau}{4}+\frac{\Delta\Lambda}{3\k^2\tau}\right)^3\right]
\right\}\ .\label{bubbleaction}
\eea
While this is a mess, the reader should note that the sign of the last term
is independent of the tension.  That is because, for $\tau<\tau_c$, 
$\sigma_+=1$ but the quantity in the parentheses in the last term of 
(\ref{bubble1}) is the square of a negative number, so the square root 
introduces a sign.  For supercritical tension, that quantity is the square of
a positive number, but then $\sigma_+ = -1$.  We have chosen to write the
variables in this form in order to illuminate the dependence on the level
spacing.  Please see figure \ref{f:cdlbt} for the qualitative features of the
action $\Delta S$ as a function of $\Lambda_-,\Delta\Lambda$.  While in
some ways the physics depends more directly on the initial cosmological
constant $\Lambda_+$, in this paper we typically works with a fixed
final $\Lambda_-$, and $\Delta\Lambda$ depends on the same moduli that
control the bubble tension.

We should note that this action reduces to the known formulae in special
cases.  In particular, the result of CDL as quoted in KKLT,
\be\label{kkltaction}
\Delta S_E = -\frac{S_0}{\left(1+\tau_c^2/\tau^2\right)^2}\ee
is valid for all tensions when the final vacuum energy vanishes.  A related result is that, for any $\Lambda_\pm \ge 0$, as the bubble tension goes to  infinity, the decay time goes to the recurrence time of the original dS.

As final comments, let us reemphasize, following 
\cite{Coleman:1980aw,Banks:2002nm}, that the final states are not maximally
symmetric spacetimes but rather cosmological ones.  In particular, decays
with a negative final cosmological constant lead not to AdS but to a Big
Crunch singularity within the bubble.  Additionally, as mentioned in
\cite{Banks:2002nm}, these instantons are technically different from 
instanton decays of inflationary spacetimes.  It seems reasonable that for
sufficiently small decay rates that treating the initial spacetime as dS 
is a good approximation, but it remains an interesting problem to study
decays of possibly more cosmologically relevant spacetimes.

\bibliography{dsinstanton}

\end{document}